\documentclass[prl,%twocolumn,
preprint,
nofootinbib,
 amsmath,amssymb,
 aps,
]{revtex4-2}

\usepackage{graphicx}% Include figure files
\usepackage{dcolumn}% Align table columns on decimal point
\usepackage{bm}% bold math
\usepackage{float}
\begin{document}

\preprint{APS/123-QED}

\title{Transport Regimes of Underdamped Brownian Particles in a Tilted Washboard Potential}% Force line breaks with \\

\author{Trey Jiron}
\affiliation{Dept.of Physical and Environmental Sciences, Colorado Mesa University\\
 Grand Junction, CO.
}
\author{Marygrace Prinster}
\affiliation{Dept.of Mathematics and Statistics, Colorado Mesa University\\
 Grand Junction, CO.
}
\author{Jarrod Schiffbauer}
 \email{jschiffbauer@coloradomesa.edu}
\affiliation{Dept.of Physical and Environmental Sciences, Colorado Mesa University\\
 Grand Junction, CO.
}

\date{\today}% It is always \today, today,
\begin{abstract}
\begin{center}
\textbf{Abstract}
\end{center}
In this paper, a comprehensive examination of the temperature- and bias-dependent diffusion regimes of underdamped Brownian particles is presented. A temperature threshold for a transition between anomalous and normal diffusive behaviors is located, yielding a new phase diagram for the system. In the low-temperature regime, the system exhibits an apparent negative differential mobility due to persistent, long-time subdiffusion at low-bias; at high temperature (or critical bias,) the system rapidly approaches normal diffusion below an intermediate barrier height, $U_0\sim6k_BT$. By consideration of numerical results, comparison to the overdamped case, and the related Kramers multistable escape problems, it is demonstrated that the low-bias non-monotonic temperature dependence of the diffusivity, persistent subdiffusion, and negative differential mobility can be traced to inertial effects, which are evident in the oscillatory modes of the velocity power spectra at low bias. In the giant diffusion regime, the velocity power spectra exhibit coupling between the ``locked" and ``running" states, with a characteristic frequency corresponding to the principal frequency of the limit cycles of a damped, driven plane pendulum near critical bias. Non-linear second harmonic generation, corresponding to oscillatory transient anomalous diffusivity, is observed with increasing bias and decreasing temperature, further emphasizing that the low-noise diffusion problem converges to noise-free dynamics, complementing analytic results for the average velocity [L. Cheng and N.K. Yip, Physica D, 2015].
\end{abstract}

\maketitle

%\tableofcontents

\section{I. Introduction and Background}
The problem of Brownian motion in a superposition of a periodic landscape with a bias force, the so-called ``tilted washboard" potential, has been used to study a wide range of mesoscale non-equilibrium phenomena. It describes driven diffusive transport in diverse physical systems including colloids \cite{EvstigneevReimannPRE2008,VolpeAmJPhy2013,VolpePetrovPRE2008}, super-ionic conductors \cite{Fulde1975}, biomolecular motors \cite{PeskinSIAM06}, diffusion of atoms on solid surfaces \cite{FerrandoPRB92,LindenbergNJP05}, and solid-state Josephson junctions \cite{LeviQAM78}.  A range of results has been obtained demonstrating a rich phenomenology in a relatively simple system, such as diffusion enhancement \cite{ReimannPRL2001,ReimannPRE2002}, non-monotonic temperature dependence of the diffusivity \cite{LindnerSokolovPRE2016,MarchenkoPRE2018,SpiechowiczPRE20}, and in some cases, regions of negative differential mobility, i.e. a local minimum in the diffusivity with respect to the bias force \cite{BerezhovskiiDagdugJCP2019}. Such effects are of more than purely fundamental interest, having applications in fields such as biomolecular sensing and separations~\cite{SlapikPRL2019}, colloidal assembly and active materials~\cite{Wang2019, Lowen2020, Petrov2021, LowenPRL2019, GaneshPRE2020}.\\
\indent Starting with Stratonovich \cite{Stratonovich}, the existence of tilt-dependent regimes of ``running" vs. ``locked" solutions suggested that such systems may be described by a two-state Markov process. Extensive theoretical studies have since been carried out of the Fokker-Planck equation governing diffusion of probability for the whole ensemble \cite{LangBook, RiskBook,JungRiskenZFPB1984}. From the point of view of Langevin equations for the over-damped system, Reimann et al \cite{ReimannPRL2001, ReimannPRE2002} demonstrated giant diffusion enhancement is a universal effect, and furthermore that the enhanced diffusion peak becomes infinitely high and narrow in the limit $T\rightarrow 0$ for the overdamped case. Using a modified Lifson-Jackson formula for the overdamped case, Berezhkovskii and Dagdug~\cite{BerezhovskiiDagdugJCP2019}, found that the shape of the potential controls the effective diffusion, resulting in a single maximum (for sinusoidal potentials), a local minimum prior to an asymptotic approach to saturated enhancement (for ``cusped" potentials,) and a monotonic approach to saturation (for square-well potentials.) Lindner and Sokolov~\cite{LindnerSokolovPRE2016} studied the temperature dependence in the underdamped case and demonstrated a non-monotonic temperature-dependence of the enhanced diffusion for a finite range of bias forces. Similar effects are seen for time-periodic forcing in spatial potentials~\cite{MarchenkoPRE2018}. This is explored in detail in a recent paper by Spiechowicz and Luczka~\cite{SpiechowiczPRE20} in the underdamped case, where it is argued that the non-monotonic temperature dependence arises due to a temperature-driven change evident in the long-time stationary velocity probability distribution.\\
\indent While Brownian motion in a tilted periodic potential is well-studied, there is little unifying or clarifying discussion of the dynamics in these different regimes. Here, we aim to remedy this by recourse to detailed numerical studies, enabling their analysis using the power-spectra of the velocity to gain more insight into the dynamics. By comparison to the zero-noise power spectrum, the results demonstrate numerically the convergence of the full, time-dependent dynamics towards those of the noise-free case. This complements an important analytical result; the long-time average velocities and transition times in the underdamped tilted periodic potential have been shown to converge to those of the noise-free system in the limit of low noise by Cheng and Yip~\cite{ChengYipPhysD2015} at moderate to high bias, i.e. in the region near the giant-diffusion effect.\\
\indent The remainder of the paper is structured as follows: In Section II the chosen normalization and numerical implementation of the Langevin equation are explained. In Section III, initial simulation results and qualitative observations are introduced for particle diffusivity over a range of bias force along with a new phase diagram in well-depth vs. temperature. To connect the range of diffusive behaviors to the underlying dynamics, the zero-bias dynamics, jump dynamics, and zero-noise system are discussed separately before considering the role of bias across temperature on the full dynamics of the system. Section IV then reiterates and discusses the main conclusions, along with possible applications and future work.\\
\section{II. Model and simulation}
We employ a standard Langevin equation for a single Brownian particle of mass $m$ immersed in a bath at uniform temperature $T$, with the particle subject to a biased cosine potential, e.g. the ``tilted washboard" potential $U(x)=U_o\cos{(\kappa x)} +F_ox$,
\begin{equation}m\ddot{x}+\gamma\dot{x}-\kappa U_o\sin{\kappa x} + F_o -\sqrt{2\gamma k_BT}\xi(t)=0\label{eqn:e1}\end{equation}
where $\gamma$ is the viscous friction of the bath and the noise term obeys $\langle\xi(t)\rangle=0$ and $\langle\xi(t)\xi(0)\rangle=\hat{\delta}(t)$. The length is scaled by the inverse wavenumber $\kappa^{-1}$ and the time by the viscous relaxation time, $m/\gamma$, so that the governing equation becomes
\begin{equation}\ddot{y}+\dot{y}-\alpha\sin{y} +\beta -\delta\eta(t)=0\end{equation}
with a dimensionless Gaussian noise $\eta(t)$.\\
\indent The three dimensionless parameters are the coupling to the bath, $\delta=\kappa\sqrt{2m k_B T}/\gamma$; the external bias, $\beta=m\kappa F_o/\gamma^2$; and the potential landscape, $\alpha= m\kappa^2 U_o/\gamma^2$. It will be helpful to the reader to note that $\delta\sim{T^{1/2}}$. These three dimensionless numbers represent ratios of the characteristic energies of respective processes in the system to viscous dissipation. Here, for a temperature-independent friction, the scaling places us in the intermediate-high-damping regime of the corresponding Kramers problem.\\
\indent Equation~\ref{eqn:e1} is reduced to a pair of coupled, first-order equations by defining $\dot{y}=v$ and explicit trajectories $y(t)$ are obtained numerically using RK4 integration implemented in Python. Due to the possible long run times of each trajectory and the large number of trajectories required for any meaningful statistical information to be extracted, efficient calculation requires parallel processing on an Intel Xeon ``Haswell" cluster. The effective diffusivity is then computed directly from the variance of the trajectories,
\begin{equation}D=\lim_{t \to \infty} \frac{\langle y^2(t)\rangle-\langle y(t)\rangle^2}{2t}\label{eqn:e2}\end{equation}
with ensembles of $N=1000-2000$ trajectories run for $10^6$ to $10^7$ time-steps and total system times of $10^4$ to $10^5$ (dimensionless time, normalized by the viscous relaxation time) averaging over every $10^{th}$ step of the last $10^4$ points in each trajectory. A well-defined constant diffusivity does not exist for all parameter ranges; there are regimes of persistent, anomalous diffusion \cite{SpiechowiczPRE20}. Thus, it is convenient to introduce the time-dependent diffusion coefficient,
\begin{equation} D(t)=\frac{\langle y(t)\rangle^2-\langle y^2(t) \rangle}{2t}\label{eqn:e5}\end{equation}
which will facilitate our discussion in super- and sub-diffusive regimes.\\
\section{III. Results and Discussion}

\begin{figure}[H]
\centering
      \includegraphics[width=3.25 in]{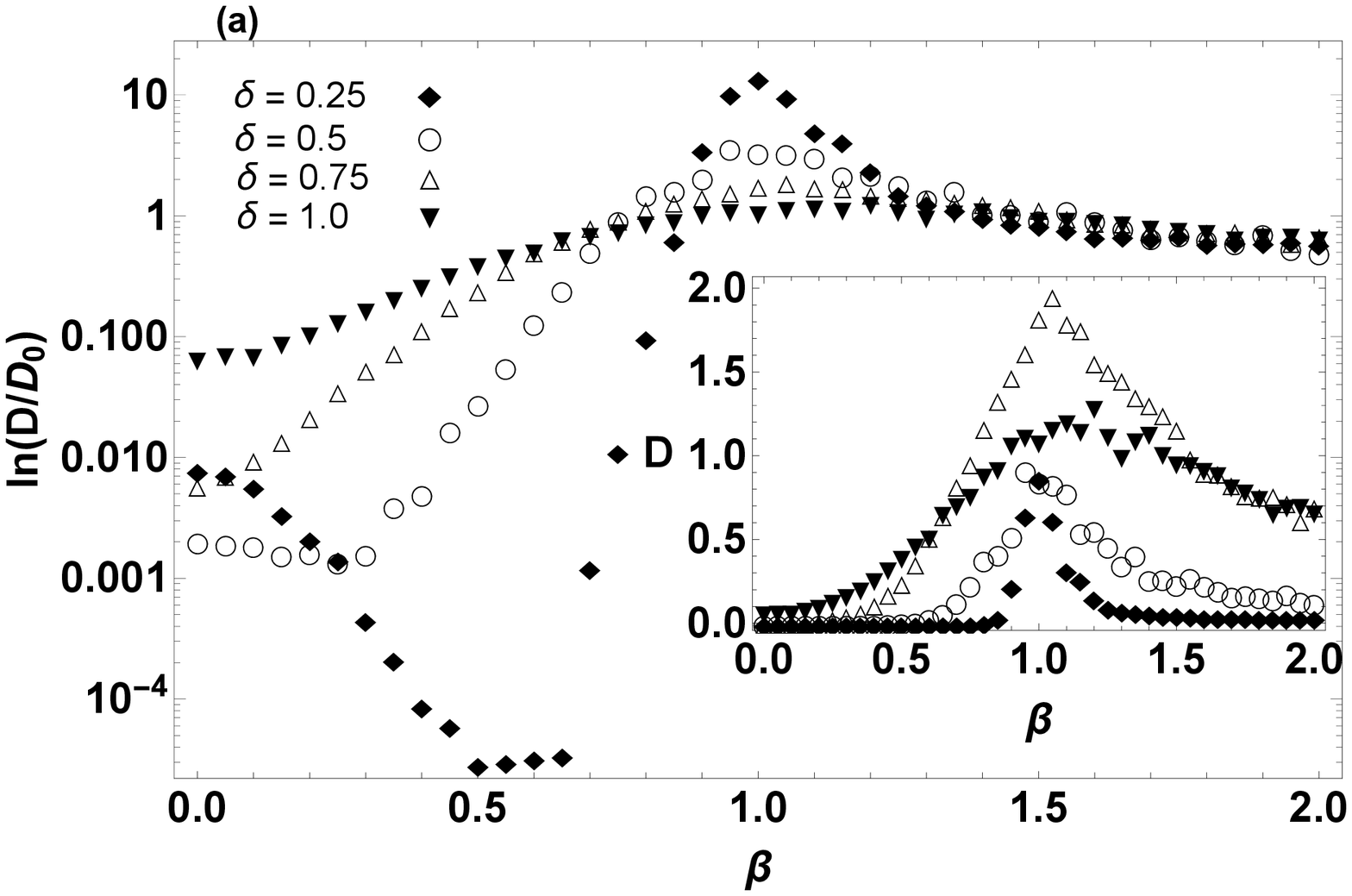}
      \includegraphics[width=3.25 in]{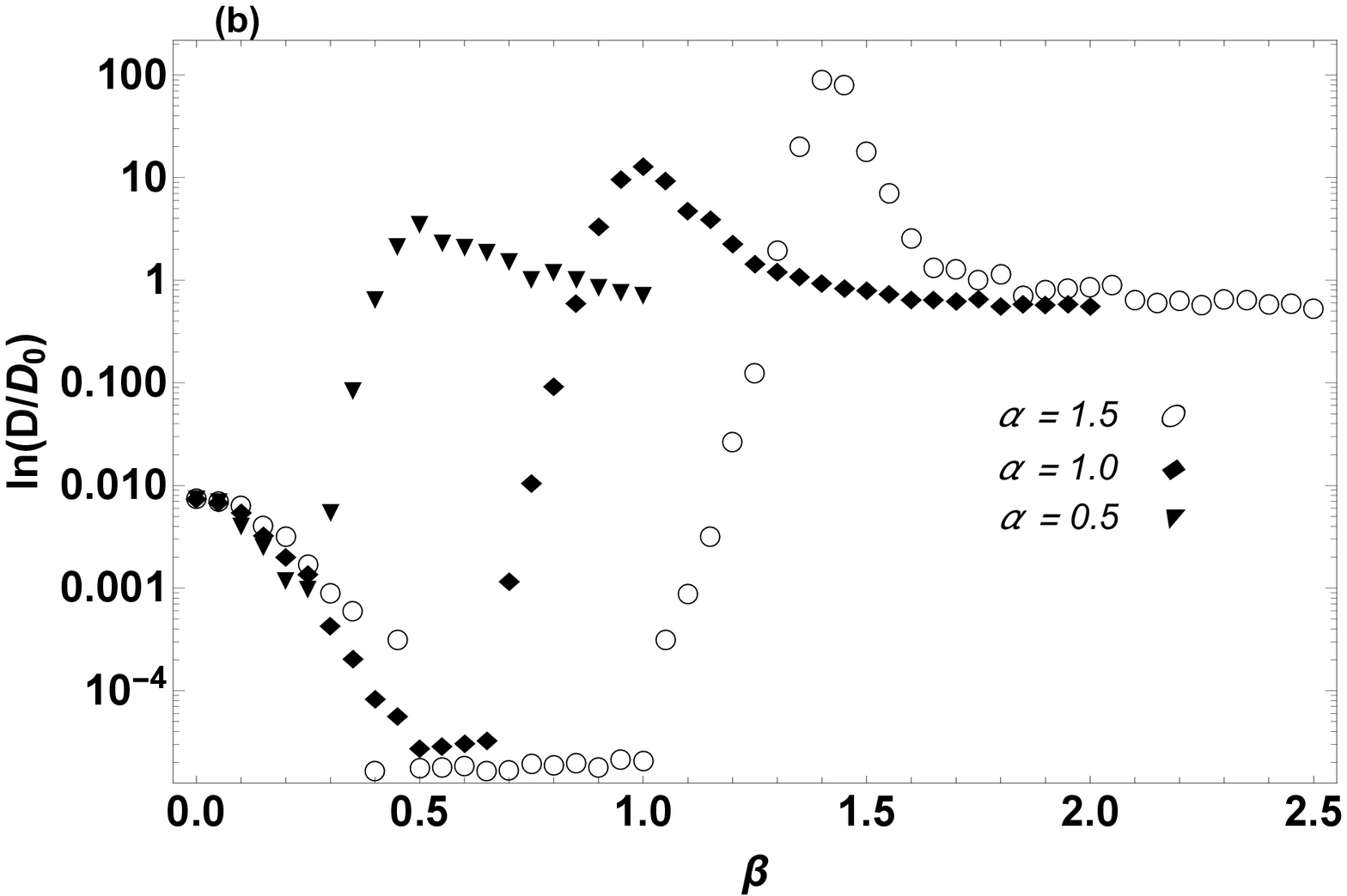}
  \caption{Effective diffusivity vs. bias parameter, $\beta$, normalized by Einstein-Sutherland value. Top (a) figure shows curves for $\alpha = 1$ and a range of $\delta$. Inset shows un-normalized diffusivity in neighborhood of giant diffusion maximum. Note that negative differential mobility is only evident in the log scaled y-axis. Bottom (b) figure shows curves at $\delta = 0.25$ for several values of $\alpha$.}\label{fig:f1}
\end{figure}
\begin{figure}[H]
\centering
  \includegraphics[width=3.25 in]{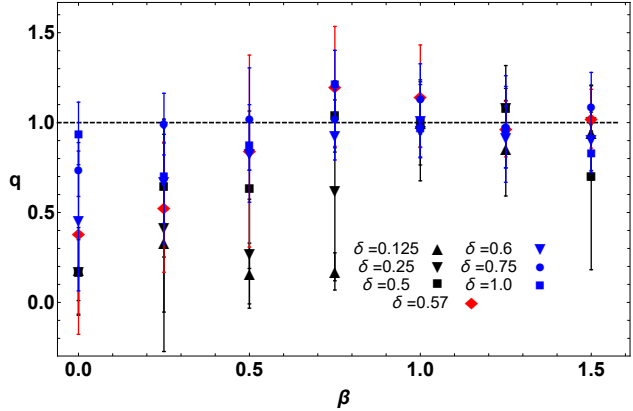}\\
  \caption{Fitted exponents from mean square deviation of trajectories at long times. The black states lie below the transition $\delta_c \sim 0.57$ (denoted by the red markers) for $\alpha = 1$ and a range of bias; blue states lie above.}\label{fig:f2}
\end{figure}
\begin{figure}[H]
\centering
  \includegraphics[width=3.25 in]{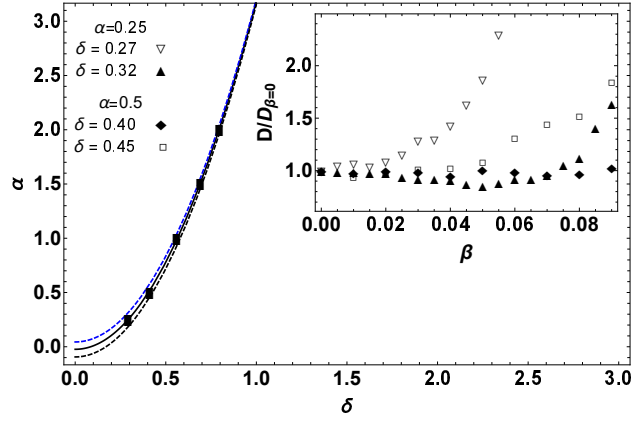}\\
  \caption{Phase diagram in $\alpha-\delta$ plane. The dashed curves correspond to $m= 2.25$ (black) and $m= 1.87$ (blue); the solid black curve is the fit for $m=2$. The region above the curve(s) corresponds to the intermediate-high damping regime in terms of the Kramers turnover. Inset shows diffusivity normalized by zero-bias value used to obtain bracketing values of $\delta$ for $\alpha = 0.25, 0.5$. Brackets for the points at $\alpha = 1.0, 1.5, 2.0$ were similarly obtained.}\label{fig:f3}
\end{figure}

The effective diffusivity vs. bias, $\beta$ is shown in Figs.~\ref{fig:f1}. The top figure shows the diffusivity for $\alpha = 1.0$ and several values of $\delta$, as indicated. Let us make a few qualitative observations regarding the response of the diffusion coefficient. The well-known giant diffusion effect~\cite{ReimannPRE2002} and non-monotonic temperature response of the giant diffusion~\cite{LindnerSokolovPRE2016} (emphasized in the inset) are evident. Another curious response is also evident, as emphasized by the vertical log scale; below the onset of giant diffusion, the effective diffusivity decreases in response to increasing bias, i.e. corresponding to a negative differential mobility, and then increases again prior to the onset of giant diffusion. While the existence of ``locked" trajectories is well-known~\cite{Stratonovich}, this effective decrease in diffusivity with applied bias does not result in a complete or immediate vanishing of the diffusivity. Rather, at least at sufficiently low $\delta$ ($\sim T^{1/2}$), there is a slow approach to a minimum diffusivity, before rising again with the onset of diffusion enhancement. A non-monotonic minimum for cusped potentials has been reported in the overdamped case~\cite{BerezhovskiiDagdugJCP2019}, the phenomenon has not been previously reported for the cosine potential in either underdamped or overdamped regimes. This effect has a pronounced temperature dependence, in particular, disappearing above some approximate temperature threshold. Similar phenomena occur in the transition from static-to-kinetic friction, grain boundary hysteresis in shape memory alloys and other stick-slip phenomena~\cite{Bhattacharya99}. At high bias, the diffusivity is dominated by the frictional response, with a weak temperature dependence (for temperature-independent friction.) What is noteworthy is that the width of the transition from a region of apparent minimum effective diffusivity through the giant diffusion maximum displays similar behavior for decreasing $\delta$ as it does for increasing $\alpha$. As $\delta$ decreases or $\alpha$ increases, there is a widening and lowering of the apparent “locked” plateau of low but non-zero diffusivity, as well as a “sharpening” and (finite) increase of the diffusion maximum peak.\\
\indent To better characterize the nature of the diffusion for various parameter ranges, the exponent, $q$, in the mean-square deviation at long times, e.g. $MSD\sim t^q$ is calculated. As usual, $q\sim 1$ denotes normal diffusion, whereas $q<1$ denotes sub-diffusive and $q>1$ super-diffusive relaxation (with the ballistic limit, $q=2$.) To do this, we sample the MSD at 4 time-intervals of increasing length, i.e. the last $n\Delta t=1000n$ times up to $t=10,000$, and display the average of fitted exponents, along with error bars indicating standard deviation in the exponents across time intervals, for a range of $\delta, \beta$ values in Fig.~\ref{fig:f2}.\\
\indent Several interesting observations can be made from Fig.~\ref{fig:f2}. First, overall, the states above the rough threshold range of $\delta \sim 0.5-0.57$ for $\alpha = 1$ generally reach normal diffusion by the last interval of $\Delta t = 1000$ for all bias values, while those below only do so for biases corresponding to diffusive amplification (and higher). Second, for non-zero bias across all parameter ranges below the onset of diffusive amplification, i.e. $0<\beta < \beta_c$ for some $\beta_c$, there is some degree of persistent, long-time sub-diffusion due to the bias itself, corresponding to the apparent negative differential mobility. As the bias approaches a critical value, here $\beta \sim 1$, the system rapidly approaches normal diffusion for $\delta$ in the range $0.125$ to $1.0$, with a marked reduction in standard deviation about $q=1$ as compared to sub-critical forcing.\\
\indent The above results demonstrate that the overall behavior of diffusion coefficient as a function of $\beta$ changes as $\delta$ increases. Corresponding to the transition from a largely subdiffusive regime to one of essentially normal diffusion, we note that the slope of the diffusion coefficient with respect to $\beta$ changes from negative to positive around $\delta\sim0.5$. By examining the decrease in diffusivity with temperature ($\delta$) for fixed well-depth ($\alpha$), as exhibited in Fig.~\ref{fig:f1}a and b, it is clear that the apparent negative differential mobility occurs when the well depth becomes sufficiently large compared to the thermal energy. Normalizing the diffusion coefficient by the zero-bias value (see inset of Fig.~\ref{fig:f3}), the transition occurs about the curve shown in Fig.~\ref{fig:f3}, which can be anticipated to go as $\delta^2$.\\
\indent We verify our intuition by fitting the high and low points bracketing the transition to a power-law curve,
\begin{equation} \alpha = \lambda \delta^m\label{eqn:e4}\end{equation}
where the exponent $m$ is found to lie between $1.87$ and $2.25$ (corresponding to the dashed curves in Fig.~\ref{fig:f3}.) Proceeding to assume $\alpha=\lambda \delta^2 +\alpha_o$, we find $\lambda = 3.209$ for the middle curve, which indicates that, roughly speaking, $U_o \sim 6k_B T$. Note that this temperature-dependent transition in diffusivity is evident in the zero-bias diffusivity of Fig.~\ref{fig:f1}a. As $\delta$ increases, the lowest diffusivity in the plot occurs near $\delta = 0.5$, with the value rising back up as the temperature increases. \\
\subsection{Zero-bias dynamics}
\begin{figure}[H]
\centering
  \includegraphics[width=3.25 in]{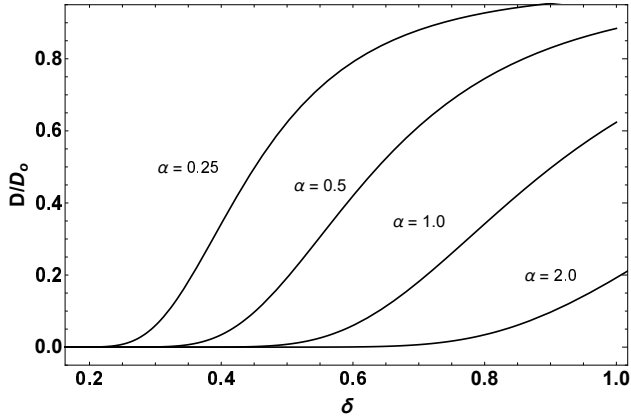}\\
  \caption{Lifson-Jackson formula diffusivity (Eqn.~\ref{eqn:e6}) vs $\delta$ for several values of $\alpha$. Rapid increase of $D_{LJ}/D_o$ shows critical $\delta$ behavior; but the overall response is monotonic}\label{fig:f4}
\end{figure}

\begin{figure}[H]
\centering
  \includegraphics[width=3.25 in]{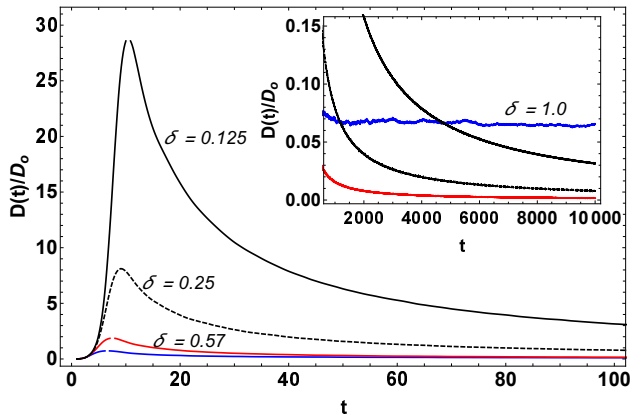}\\
  \caption{Zero-bias normalized diffusion coefficient vs. time for several temperatures (indicated) in the $\alpha=1$ case. Inset shows same figure at very long times. Lowest $\delta$ values exhibit persistent sub-diffusion at long times.}\label{fig:f5}
\end{figure}

In Fig.~\ref{fig:f1}a, it is apparent that even the zero-bias diffusivity responds non-monotonically to changes in temperature ($\sim\delta^2$). The fact that low, but non-zero bias exhibits a decrease in the diffusivity, and that the range of the non-monotonic temperature response extends from $0$ to above the critical bias, suggests that a closer examination of the zero bias response is warranted. Note both underdamped and overdamped systems exhibit a distinct change in response at zero bias with respect to $\delta$. The overdamped response can be seen in Fig.~\ref{fig:f4}, where the Lifson-Jackson formula~\cite{Lifson}
\begin{equation}\frac{D_{LJ}}{D_o}=\frac{(2\pi)^2}{\int_{-\pi}^{\pi}dy\exp{\left(-\frac{\alpha}{\delta^2}\sin{y}\right)}\int_{-\pi}^{\pi}dy\exp{\left(\frac{\alpha}{\delta^2}\sin{y}\right)}} \label{eqn:e6}\end{equation}
is used to plot the zero-bias normalized diffusivity vs. $\delta$ for $\alpha =0.25, 0.5, 1.0, 2.0$. The diffusivity varies very slowly until approaching $\alpha \sim 3\delta^2$, where the diffusivity undergoes a rapid, however monotonic, increase with temperature. Considering the un-biased periodic potential, Ferrando et al~\cite{FerrandoPRE1993} used a numerical solution of the corresponding Klein-Kramers equation, along with analytical approximations of the Melnikov type~\cite{Melnikov}, to demonstrate that the jump rates and jump length distributions had strong dependence on temperature. While this temperature-dependence (relative to barrier height) is present across a wide range of friction, the rates and probabilities for multiple jumps exhibit a distinct maximum in the underdamped regime. It is unsurprising that both overdamped and underdamped dynamics exhibit a transition around the same temperature. However, note that the overdamped case shows monotonic temperature response, in contrast to the zero-bias values shown in Fig.~\ref{fig:f1} in which $\delta\sim 0.5-0.57$ shows a minimum in zero-bias diffusivity at long time. Thus, this non-monotonic response at low bias is tied intimately to inertial effects.\\
\indent The time-dependent (zero-bias) diffusion coefficient for the underdamped case with cosine potential is plotted in Fig.~\ref{fig:f5} for several values of $\delta$. The non-monotonicity with respect to temperature is persistent at long times, consistent with the subdiffusion, and apparent negative differential mobility. For lower values of $\delta$, the diffusivity decreases with longer simulation times as the ``locked" state waiting time increases, leading to persistent subdiffusion as the noise decreases. For example, at $\tau = 100,000$, the diffusivity is about an order of magnitude smaller than at $\tau = 10,000$, suggesting that the diffusivity decreases roughly as $1/\tau$. However, as $\delta$ increases, the diffusivity settles much more rapidly (see Fig.~\ref{fig:f5}). Note the contrast between the non-monotonic temperature dependence exhibited by the underdamped system, as opposed to the monotonic temperature response of the overdamped diffusivity exhibited in Fig.~\ref{fig:f4}.\\
\indent The negative differential mobility occurs in a distinct region of the parameter space for low $\beta$ and $\delta$, corresponding to a persistent long-time sub-diffusive regime. The same temperature threshold occurs in the overdamped system, but there the zero bias diffusivity increases monotonically with temperature. Generally speaking (though not visible on the resolution of the figures), the normalized overdamped diffusivity approaches zero far faster with decreasing $\delta$ than does the zero-bias diffusivity of the underdamped system \emph{below the temperature threshold,} while at higher temperatures, the overdamped system tends to have a higher normalized diffusivity at zero bias. A deeper understanding of the situation may be obtained by considering the jump dynamics explicitly.\\
\subsection{Jump dynamics}
\begin{figure}[H]
\centering
  \includegraphics[width=3.25 in]{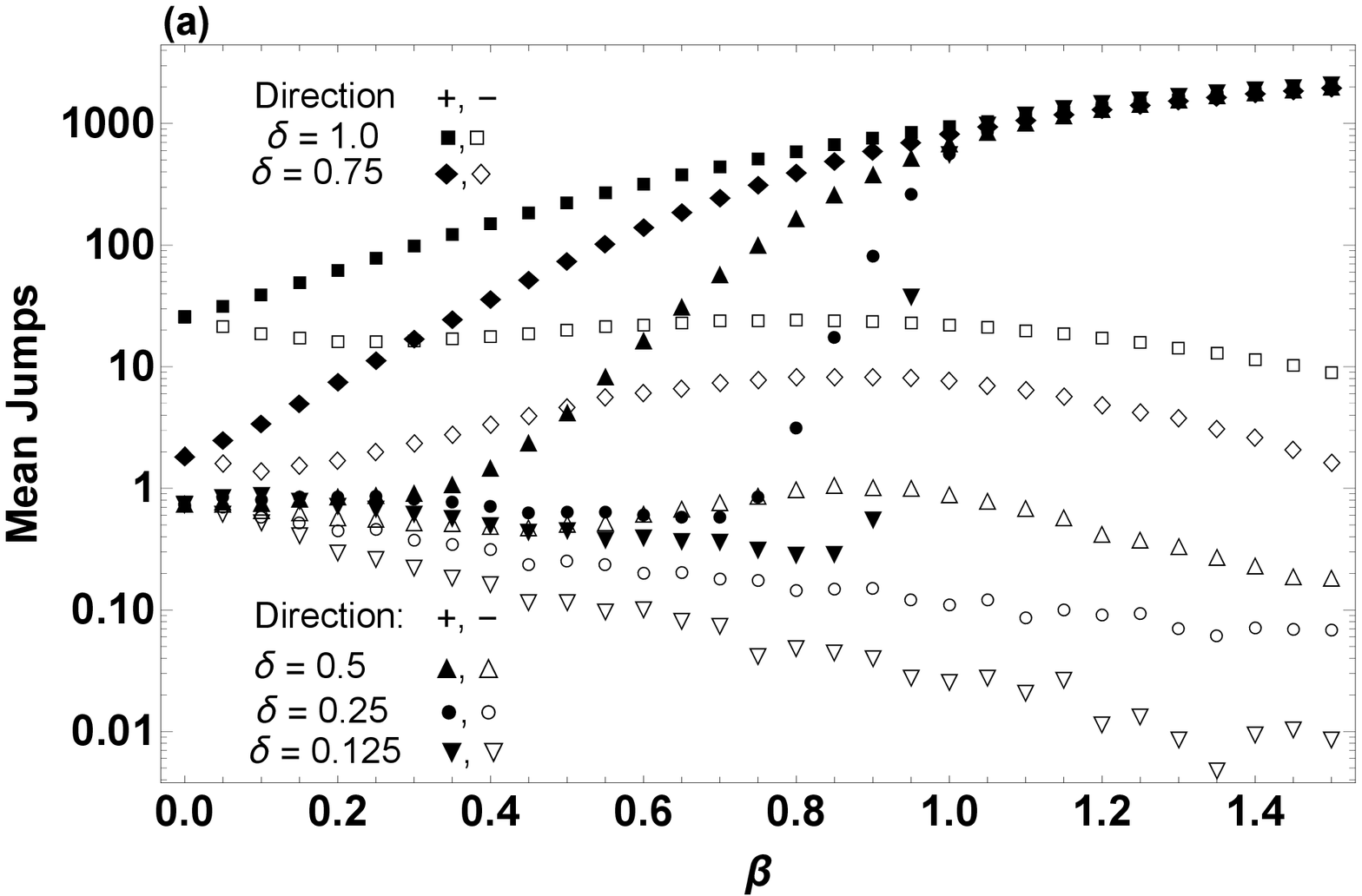}\\
  \includegraphics[width=3.25 in]{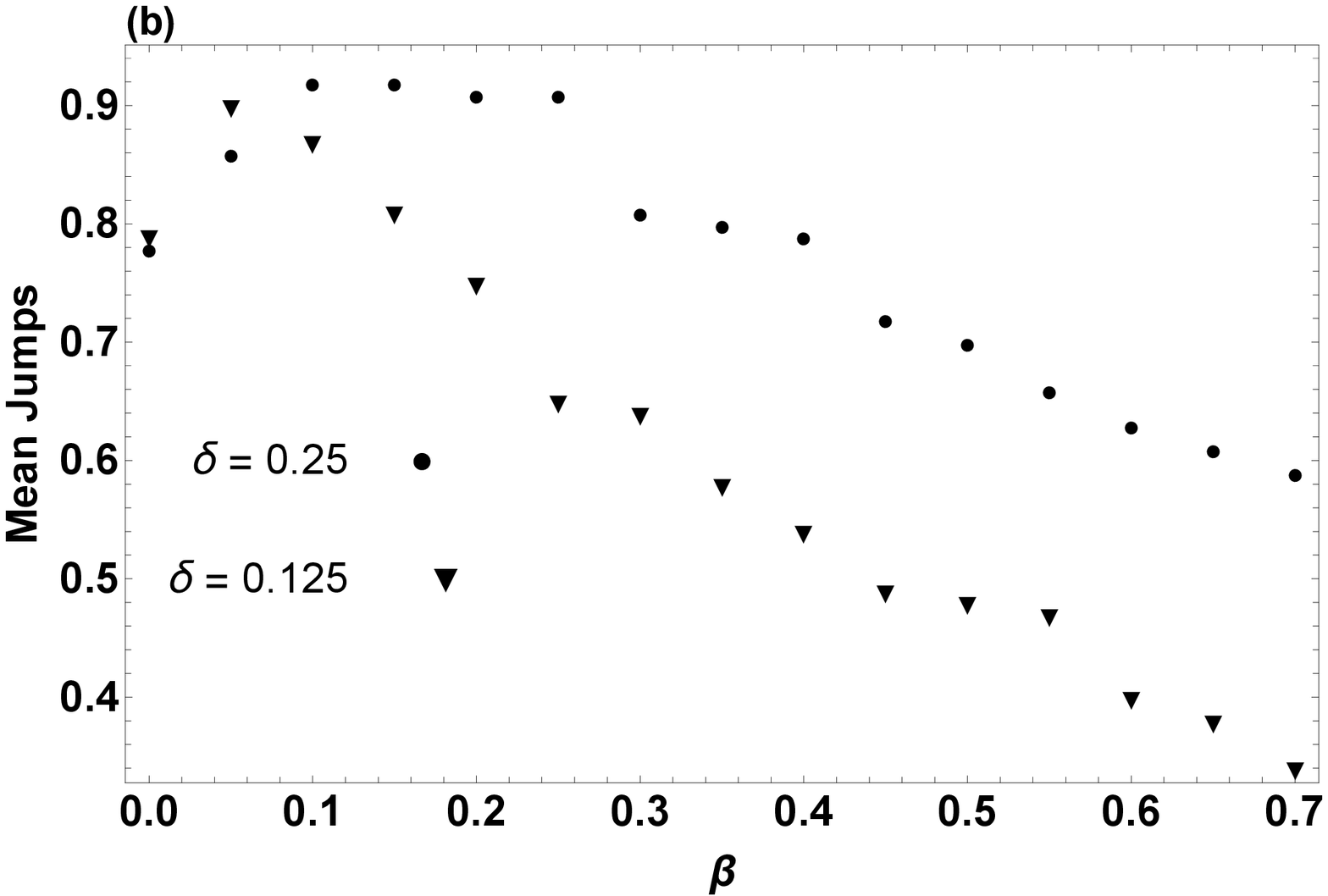}\\
  \caption{Mean directional jumps averaged over the full ensemble for $\alpha = 1$ (a) Reverse-biased (-) and forward-biased (+) jumps for a range of $\delta$  and (b) forward jumps at low $\beta$ showing the low-bias maximum at the lowest $\delta$. Since all simulations are performed for the same maximum time, $t_f=10,000$, these may be interpreted as jump rates without loss of generality (multiplying by the appropriate $10^{-4}$ factor.)}\label{fig:f6}
\end{figure}

\indent Noise-assisted escape from a periodic well involves inherently multistable dynamics\cite{FerrandoPRE1993}, and the symmetry-breaking effect of the bias introduces another complication. Our assertion is that the decrease in diffusivity with increasing bias and corresponding subdiffusion can be attributed to asymmetry in the forward and reverse-biased jump rates and a concomitant decrease in the overall jump rate at low bias. To test this assumption, the forward and reverse-biased jump rates are calculated, averaging over the well indices for forward and reverse jumps separately. \\
\indent Calculating directional well jumps from particle trajectories, naively one expects the biased system to exhibit a lower reverse-biased jump rate compared to the forward-biased rate. One also expects that the forward biased jump rate will increase as the reverse jump rate decreases. The reader will note that in Fig.~\ref{fig:f6}b, at low $\beta$ and low $\delta$, this expectation is borne out. However, it can be seen there is a maximum in the forward-biased jump rate at low $\beta$, and the corresponding value of bias for this maximum rate increases with temperature. Beyond this maximum, the forward-biased jump rate decreases to a local minimum, and corresponding with the onset of the giant diffusion effect, this rate increases dramatically. Thus, at $\delta$ below the identified transition, the forward-biased jump rate possesses both a local maximum (at very low bias,) and a local minimum immediately prior to the onset of giant diffusion. This local maximum and minimum move closer together with increasing temperature and annihilate for $\delta$ in the vicinity of $0.57$ such that the forward biased jump rate increases monotonically at higher $\delta$.\\
\indent  The reverse-biased jump rate is no less intriguing. It is worth pointing out that, above critical bias, the forward jump rate is a universal function of bias and does not depend on temperature, while the temperature-dependence of the reverse-biased jump rate is unexpected. At low temperatures, the reverse-bias rate decreases monotonically with increasing bias, as one would expect. However, nearing the transition temperature, the reverse jump rate develops a local maximum \emph{in the vicinity of the giant diffusion maximum, approaching critical $\beta$}, exceeding the zero bias reverse jump rate. This maximum increases with respect to both the global reverse jump rate and the zero bias jump rate; but critically it exceeds the zero bias value by almost an order of magnitude around $\delta = 0.75$, coincident with the maximum un-normalized giant diffusion (see inset of Fig.~\ref{fig:f1}a.) Above this, the maximum becomes less pronounced as the temperature continues to increase. As will be explained in the following section, this corresponds to bias-driven coupling between ``locked" and ``running" states across the I-III bifurcation, which broadens with respect to bias as the temperature increases.\\
\indent Keeping in mind that $\beta$ represents the characteristic work done by the external force as the particle is forced through one period, increasing $\beta$ increases the kinetic energy, bringing the lowest energy states closer to the top of the barrier and increasing the kinetic energy of those already at or above the barrier top. Eventually, at sufficiently large $\beta$ for a given $\delta$, this results in an increase in particle current more than sufficient to offset any reduction of jump rate due to asymmetry and the diffusivity begins to rise again with increasing $\beta$
\subsection{Zero-noise dynamics}
\begin{figure}[H]
\centering
  \includegraphics[width=3.25 in]{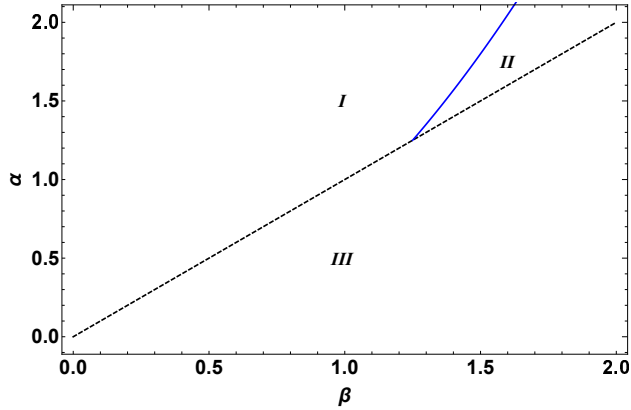}\\
  \caption{Bifurcation diagram for the noise-free dynamical system, Eqn.~\ref{eqn:e3}. This is essentially the same as Fig.(8.5.10) Strogatz~\cite{Strogatz}, but in the scaling used here. The solid blue line indicating the homoclinic bifurcation diverges from the infinite period bifurcation at $\alpha = 5/4$.}\label{fig:f7}
\end{figure}

The peculiar features of the jump dynamics, particularly the decrease in forward-biased jump rate and existence of strong reverse-biased motion, can be better understood with recourse to the behavior of the system in the absence of thermal fluctuations,
\begin{equation}\ddot{y}+\dot{y}-\alpha\sin{y} +\beta =0\label{eqn:e3}.\end{equation}
The model, equivalent to the noise-free Josephson junction or the damped, driven pendulum, has a rich oscillatory dynamics typical of inertial systems with bifurcations between three regions in Fig.~\ref{fig:f7} (see for example~\cite{Strogatz} and references therein.)  Region I contains both stable sinks and unstable saddles, corresponding to the well bottoms and tops. Region III contains limit cycle oscillations, and the Jacobian corresponding to the first-order system represented by Eqn.~\ref{eqn:e3} admits stable spirals as well as saddle nodes in Region II. The role of thermal fluctuations is particularly important around the bifurcations. Here, for $\beta$ and $\alpha$ sufficiently close, fluctuations can couple trajectories on either side of the line. This results in the giant diffusion effect itself by maximizing the spread in trajectories which initially started near the same phase point. For lower $\delta$ values, thermal fluctuations make the region III limit cycles available to phase points starting in Region I and the diffusion maximum occurs just below the $\alpha = \beta$ line. For the remainder of the paper, we confine our attention to the dynamics approaching and exceeding the I-III bifurcation. As the long-time diffusivity for $\alpha = 1.5$ (see Fig.~\ref{fig:f1}b) has similar overall behavior (at least for lower $\delta$,) we leave investigation of the dynamics at the I-II and II-III bifurcations for later study.\\

\subsection{Bias- and temperature- dependent dynamics}
\begin{figure}[H]
\centering
  \includegraphics[width=3.25 in]{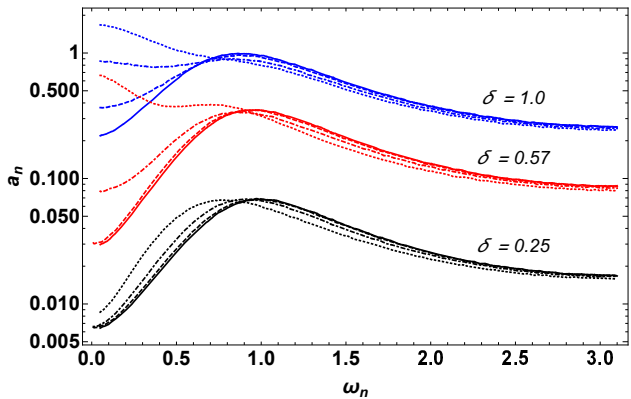}\\
  \caption{Power spectra for several cases for $\alpha = 1$ at zero bias (solid lines) and biased cases indicated by broken lines as follows: $\beta = 0.25$ (dashed), $\beta=0.5$ (dot-dashed), and $\beta = 0.75$ (dotted).}\label{fig:f8}
\end{figure}

To facilitate discussion of the role of bias by providing some insight into the time-dependent dynamics of the stochastic system, the velocity power spectrum is introduced:
\begin{equation}a(\omega)=\left|\int_{0}^{\infty}dt \exp{(-i\omega t)} v(t)\right|^2\label{eqn:e5}\end{equation}
\indent The power spectra for a range of $\delta$ and several bias values below the giant diffusion effect are shown in Fig.~\ref{fig:f8}, with solid lines indicating zero bias and biased states indicated by interrupted lines, as explained in the caption. The zero and lowest bias states exhibit a distinct maximum, with the maximum approaching the frequency of the oscillations about the well minimum as $\delta$ (temperature) increases. Note that the overall frequency distribution shifts energy to lower frequencies as the bias increases. This can be seen in the corresponding linearized damped oscillator problem, where the frequency of oscillations about the (stretched) minimum is given by the (imaginary) root $\sqrt{1-4\alpha\sqrt{1-\beta^2/\alpha^2}}/2$. Thus, below critical bias, energy shifts to lower frequencies with increasing bias. Since
$2/\delta$ gives the typical time-scale for thermal transition between wells, velocity oscillations corresponding to frequencies less than the inverse value indicate well-to-well, i.e. sink-over-saddle, transitions. Thus, as $\delta$ increases, there is more energy being made available to the low-frequency jumps, while simultaneously the thermal velocity is increasing. Still, for $\delta$ below about $0.5-0.57$ (for $\alpha=1$,) this is negligible in comparison to the energy being stored in the vibrational modes inside the wells. An important observation is that the simple-harmonic oscillatory modes in the ``locked" states arise due to the underlying oscillatory dynamics enabled by the inertial term in the zero-noise system. As $\delta$ increases above the threshold and $\beta$ increases towards $\alpha$, the power spectrum increasingly resembles the $1/f$ ``pink noise" spectrum, a characteristic of noise-driven transitions~\cite{YuPRE}.\\

\begin{figure}[H]
\centering
  \includegraphics[width=3.25 in]{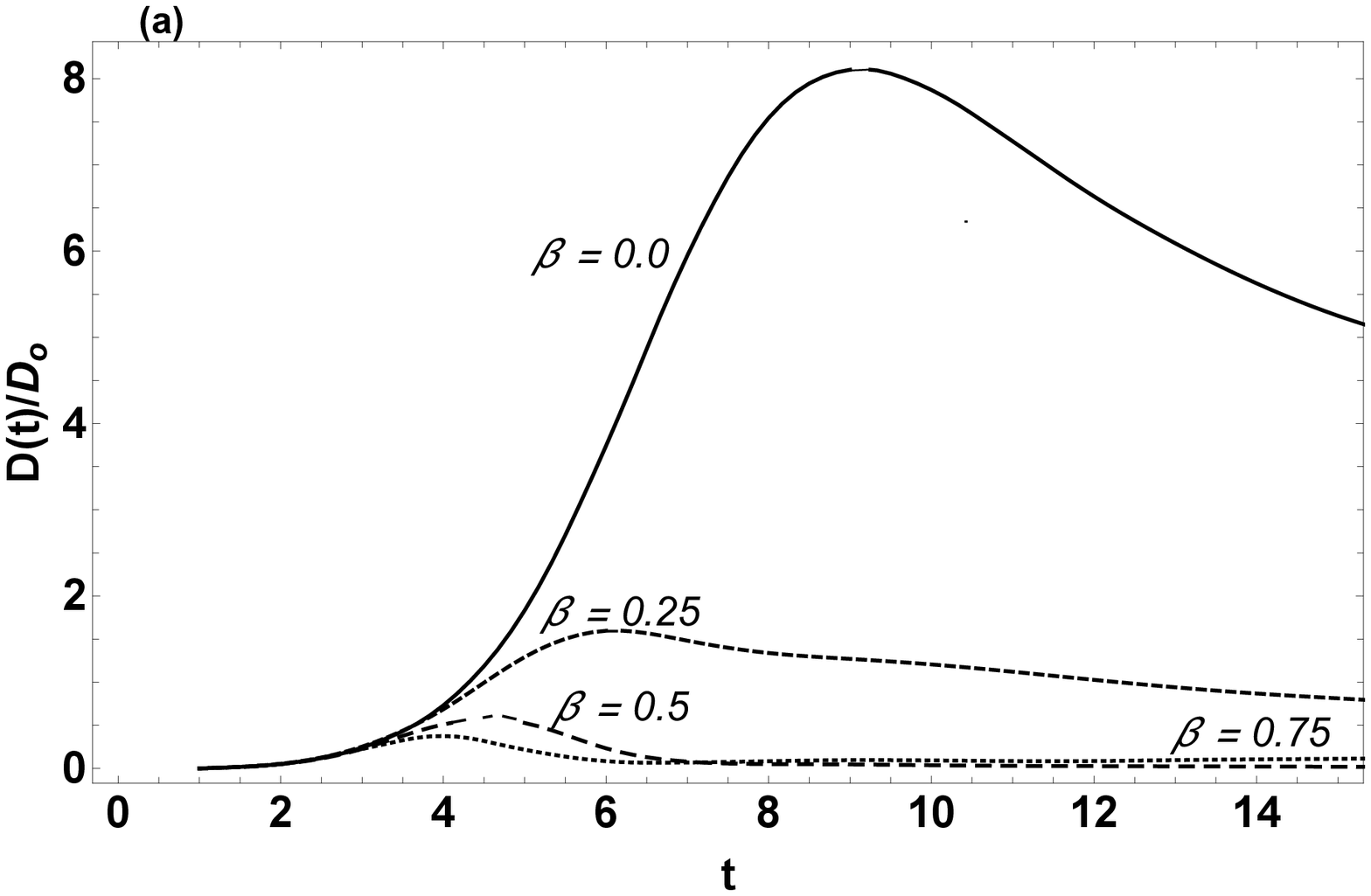}\\
    \includegraphics[width=3.25 in]{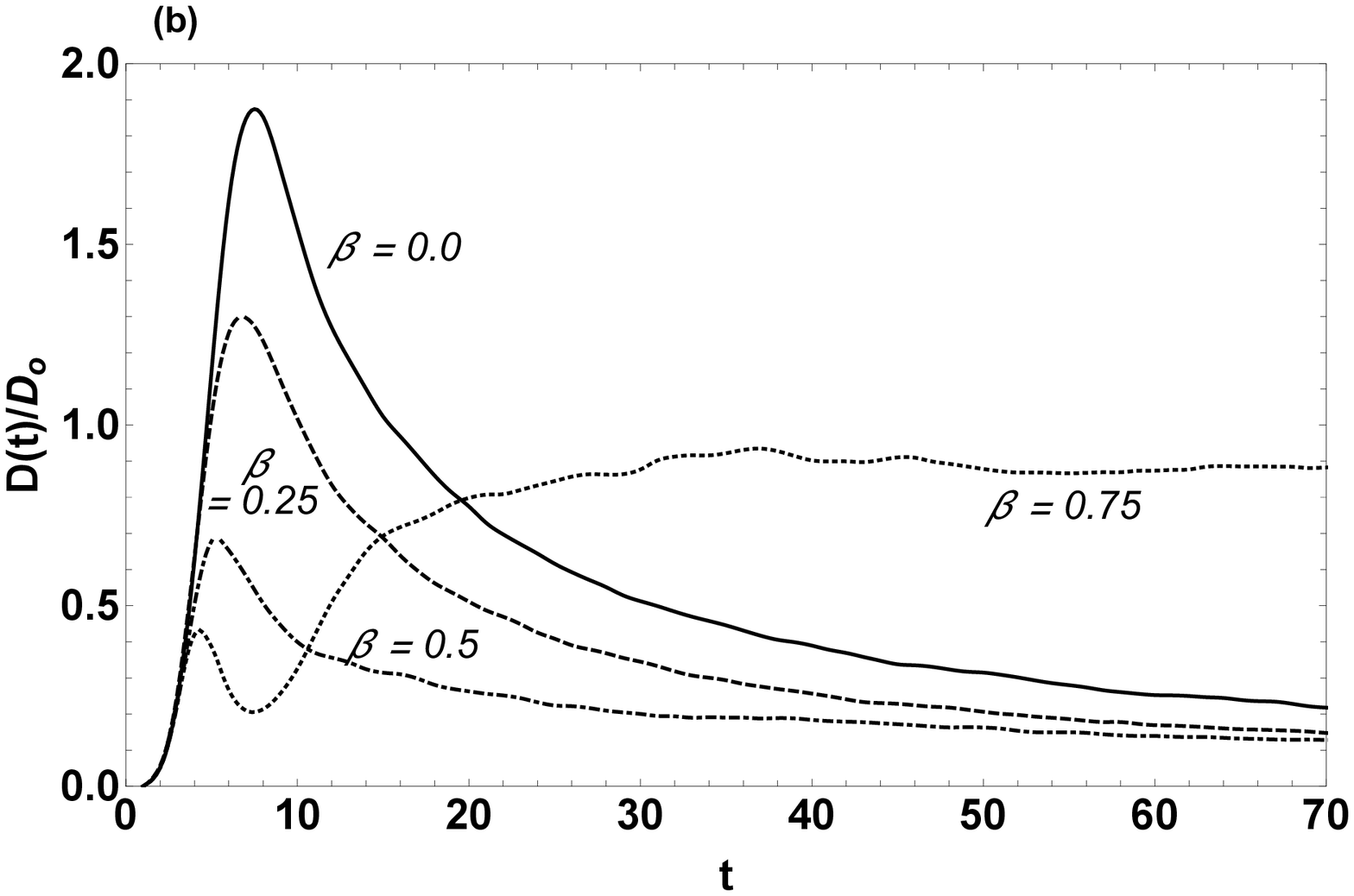}\\
      \includegraphics[width=3.25 in]{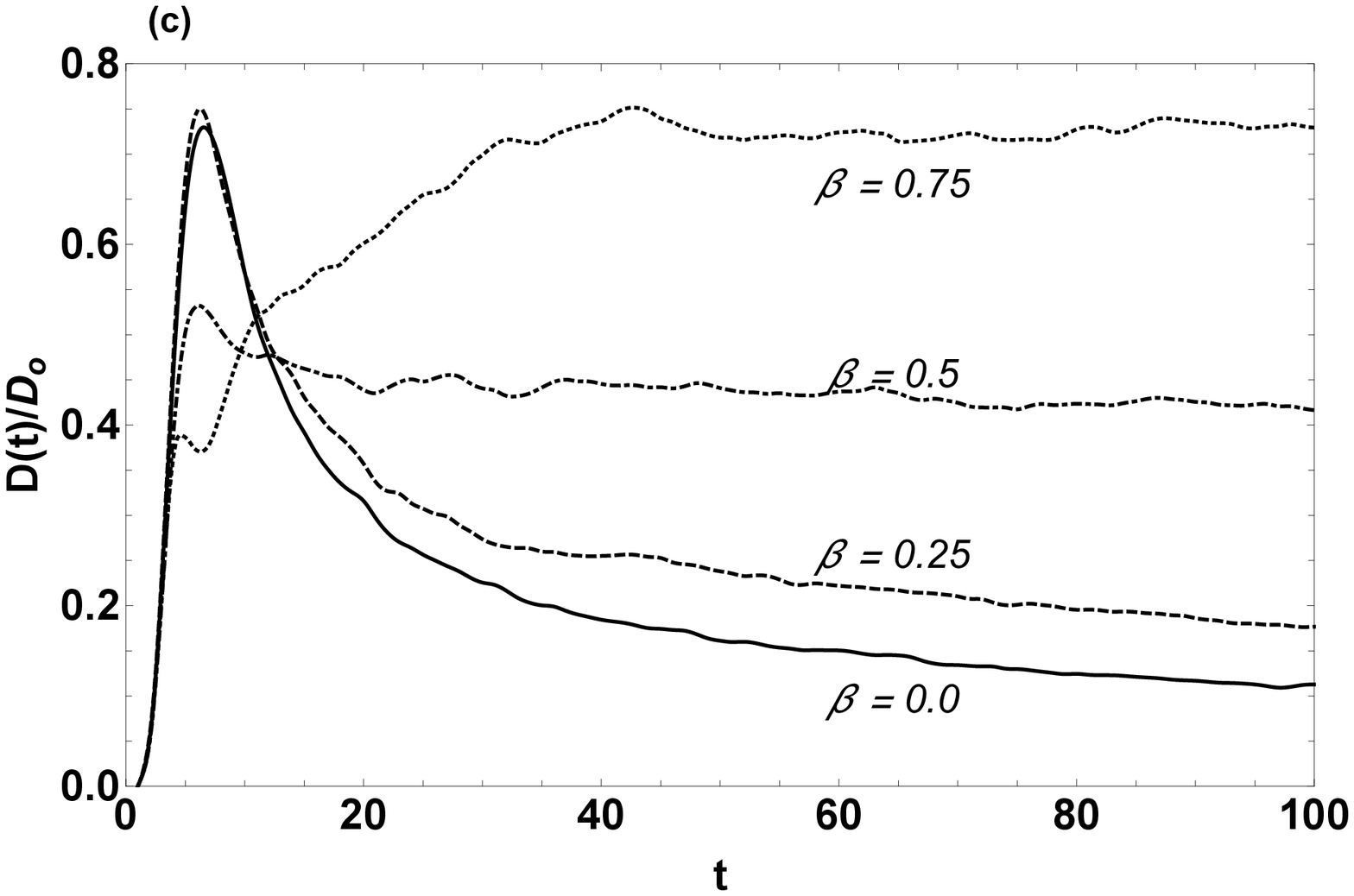}\\
  \caption{Time-dependent diffusion coefficients for several cases for $\alpha = 1$ at early times. Long-time behavior is reflected in Fig.\ref{fig:f1}. Top (a) shows decrease of initial diffusion coefficient with increasing bias for $\delta=0.25$, typical of temperatures below the critical. Figs (b) and (c) show the transition-region temperature ($\delta=0.57$)and the high-temperature ($\delta=1.0$) behavior, respectively.}\label{fig:f9}
\end{figure}
\begin{figure}[H]
\centering
  \includegraphics[width=3.25 in]{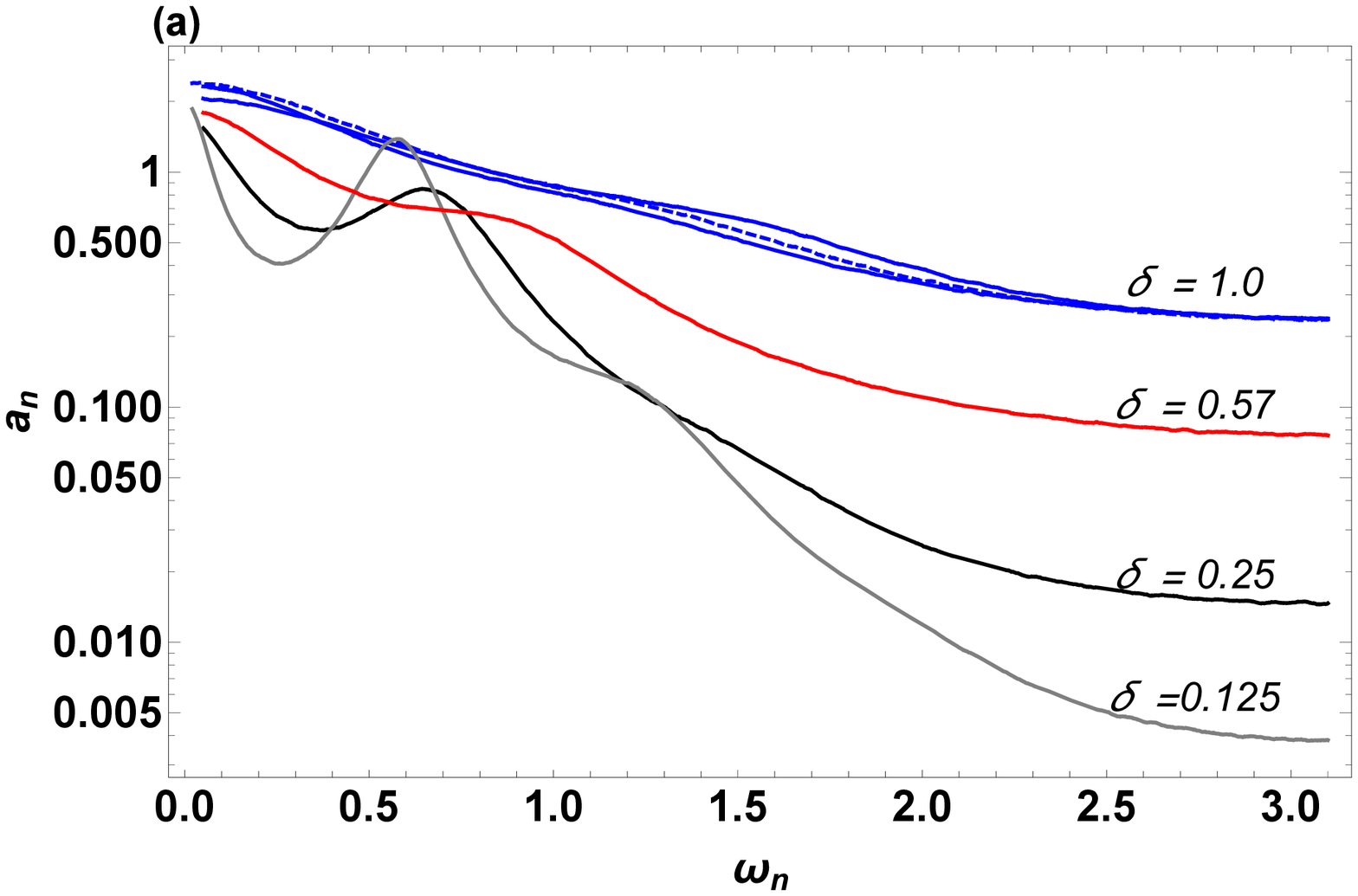}\\
    \includegraphics[width=3.25 in]{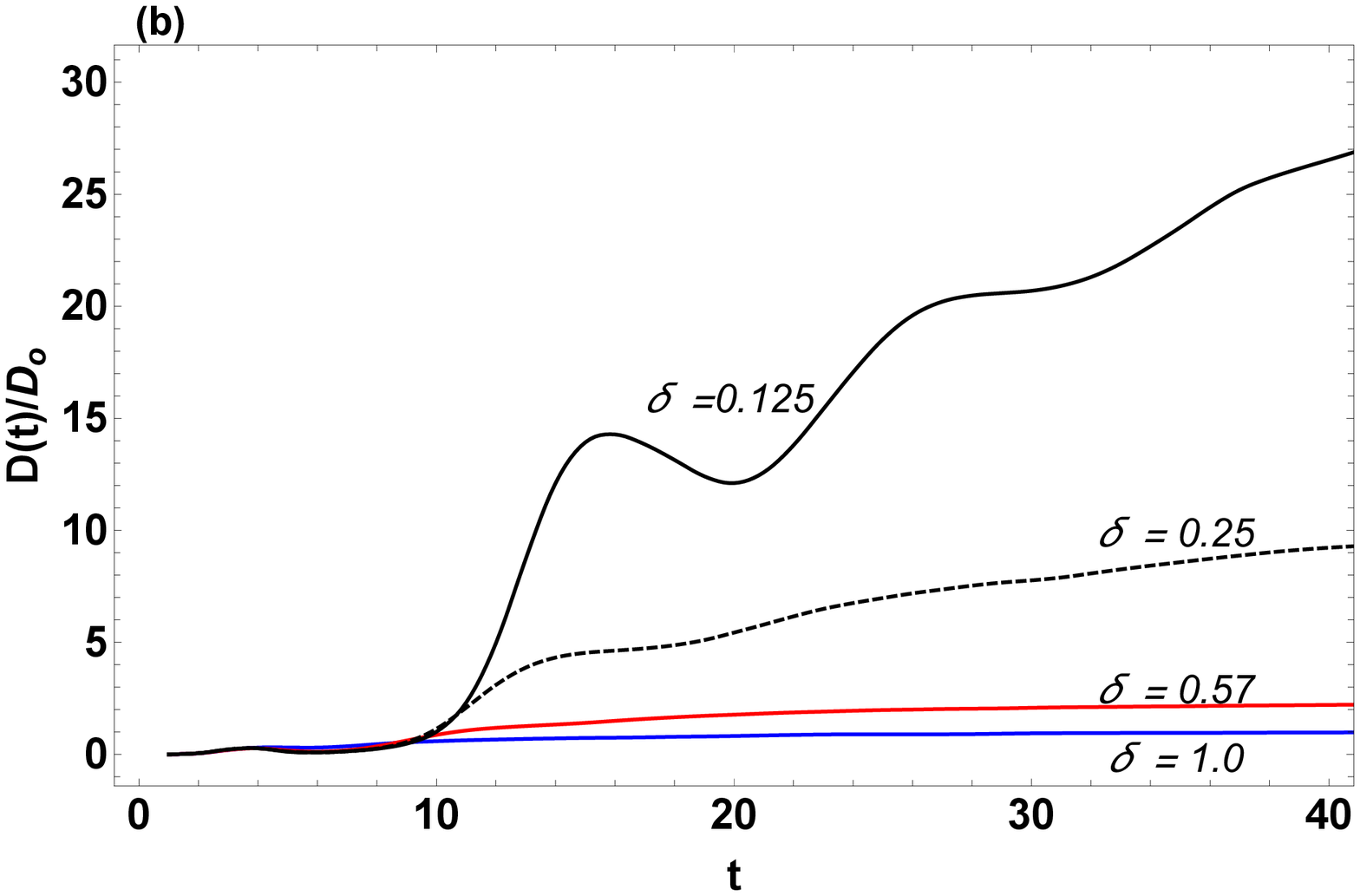}\\
  \caption{(a) Power spectra and  (b) diffusion coefficient for several temperatures near critical bias, i.e. the bifurcation line $\alpha=\beta=1$. The top curves in the power spectra (for $\delta=1$) are for $\beta = 1,1.25,1.5$ (solid, dotted, dashed, respectively) and show little variation from a $1/f$ ``pink noise" spectrum. }\label{fig:f10}
\end{figure}

\indent A comparison of Fig.~\ref{fig:f5} and Fig.~\ref{fig:f9} shows that the increase in tilt and increase in temperature can both result in reduced peak anomalous diffusivity and faster subdiffusive decay to normal diffusion. The peak diffusion coefficient at the end of the ballistic-dynamical transient ($t\sim10$) decreases with increasing bias for all cases. But the effect is more drastic at lower temperatures.\\
\indent The diffusion coefficients and power spectra as $\beta$ approaches $\alpha$ in an infinite-period (I-III) bifurcation are shown for several values of $\delta$ in Fig.~\ref{fig:f10}. As $\delta$ increases, the power spectrum approaches $1/f$, as mentioned above. However, below $\delta \sim 0.57$, a distinct peak emerges from a flat band of frequencies with decreasing $\delta$. At the lowest $\delta$ values shown, these peaks indicate the shift of energy to intra-well oscillations as temperature decreases for bias near critical. This manifests as a ``ringing" in the early time-dependent diffusion reminiscent of the transient, decaying oscillations of the damped linear harmonic oscillator subject to an impulsive load. Here, however, the frequency corresponds to region III non-linear limit cycle oscillations thermally coupled to the sinks and saddles of region I.\\
\begin{figure}[H]
\centering
  \includegraphics[width=3.25 in]{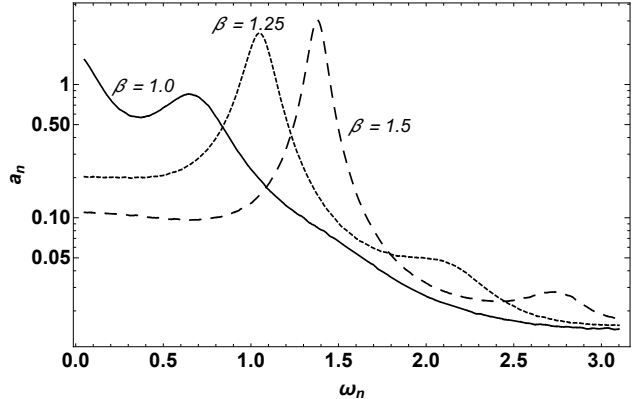}\\
  \caption{Power spectra for $\delta = 0.25$ ($\alpha = 1$) case at several beta showing state just above max ($\beta = 1$) through $\beta = 1.5$, where second harmonic generation is evident}\label{fig:f11}
\end{figure}
\indent In Fig.~\ref{fig:f11} frequency peaks increase in height with increasing bias, demonstrating a shift of energy to intra-well oscillations with bias exceeding critical at low temperature. As Figure 12 demonstrates, corresponding peaks for the zero-noise system can be expected to come from frequencies approximately equal to the nondimensional bias value. However, it is notable in Fig.~\ref{fig:f11} that peak frequencies at low noise seem to shift lower than their corresponding zero-bias values with decreasing bias. At $\beta>\alpha$, the peak frequencies correspond very well to zero-noise peaks, and the non-linear second harmonic emerges from the flat band of frequencies above this. The power spectra for $\beta = 1.5$ is shown in Fig.~\ref{fig:f12} along with the time-dependent diffusion coefficient for a range of $\delta$. The zero-noise power spectrum is also shown, where the dynamics clearly appear to converge as $\delta\rightarrow 0$. The limit cycle dynamics of the damped, driven pendulum provides a useful, mechanistic understanding of the nature of the ``running" states. Importantly, the velocity has periodic fluctuations about a well-defined average, with multiple time-scales associated with typical non-linear oscillations. These can be seen in Figs.~\ref{fig:f11} and~\ref{fig:f12}. However, the distinctions between the meanings of the coordinates aside, the Brownian system cannot follow true limit cycles, as the trajectories are nowhere differentiable, and hence not unique. This allows for coupling bands of nearby trajectories via thermally-induced transitions, hence broader peaks and shoulders in comparison to the sharper spectrum of zero-noise dynamics (see Fig.~\ref{fig:f12}.)\\
\indent This suggests, rather than a phase space containing distinct multi-stable ``running" velocity states, the topology more closely resembles that of the damped, driven pendulum (taking obvious liberty with the meaning of $y(t)$.) The diffusion coefficient displays strong oscillatory ringing at the lowest $\delta$, which is, in principle, experimentally observable in the mean-square deviation at early times for a system with a very high barrier, $U_o\sim32k_BT$ and larger for $\alpha=1$, driven above giant diffusion.
\begin{figure}[H]
\centering
  \includegraphics[width=3.25 in]{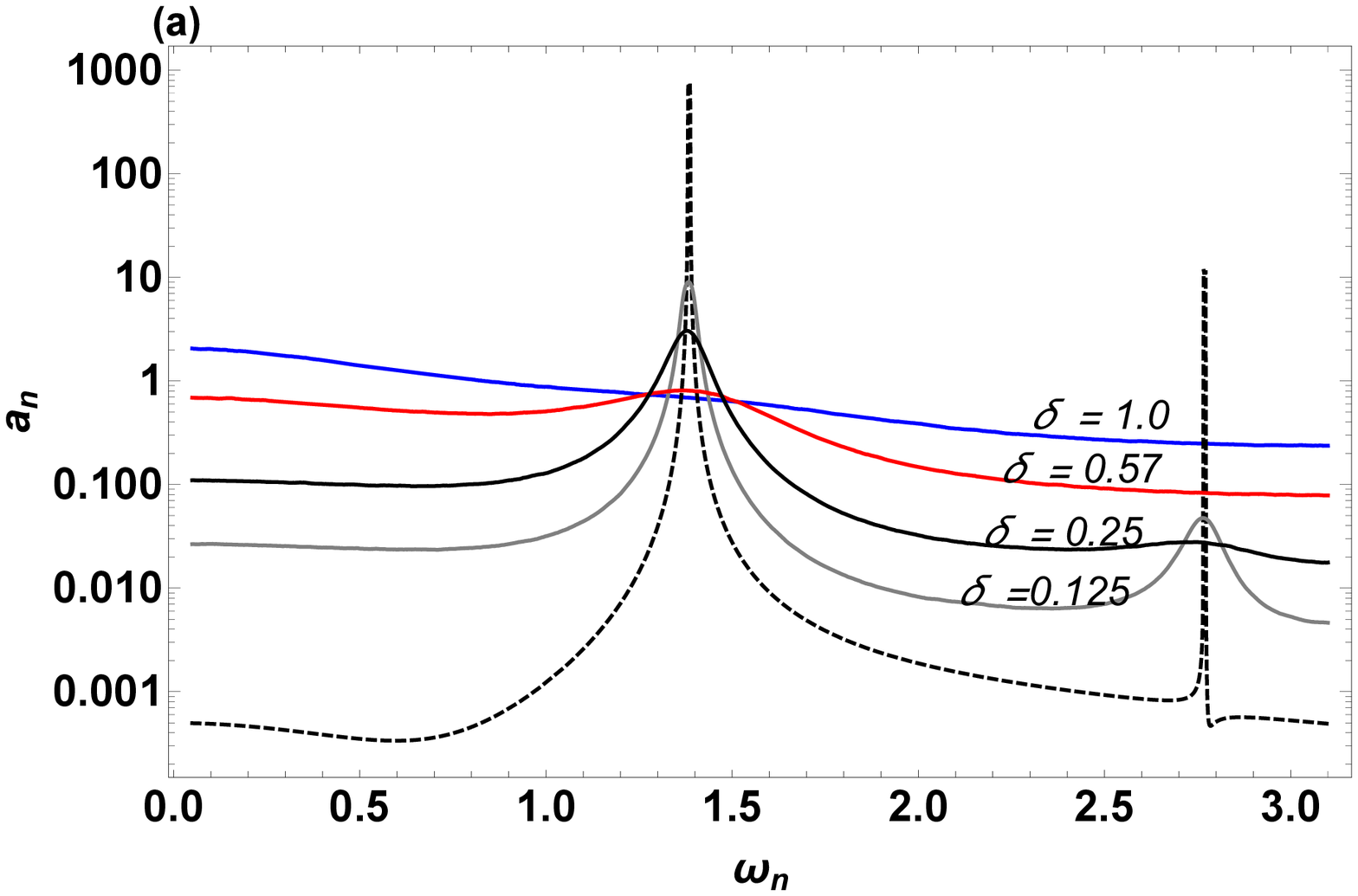}\\
    \includegraphics[width=3.25 in]{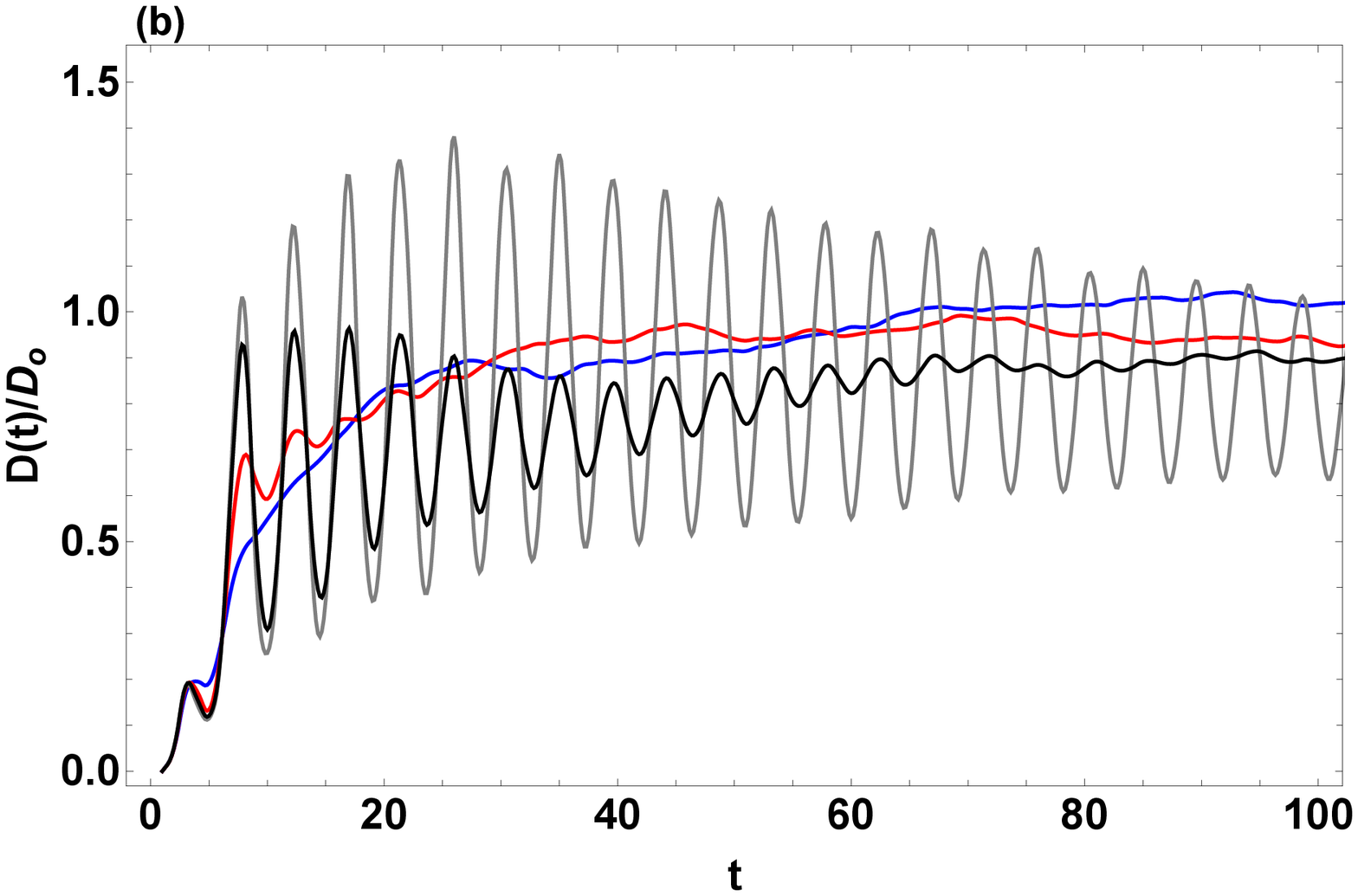}\\
  \caption{(a) Power spectra with zero-noise case shown (dashed) and (b) time-dependent diffusion coefficient for $\alpha=1$, $\beta = 1.5$, and several $\delta$ showing onset of bifurcation/node coalescence and subsequent second harmonic generation.}\label{fig:f12}
\end{figure}
\section{IV. Discussion and Conclusions}
\indent In this paper, we have revisited the classic problem of underdamped Brownian motion in a tilted washboard potential via extensive numerical simulations, complementing our discussion with a review of the corresponding zero-noise dynamics and the related mulitstable Kramers problem. A picture emerges of stochastic dynamics in a complex three-dimensional parameter space, where now, in addition to the previously known friction vs. bias phase diagram~\cite{SpiechowiczPRE20}, a well-depth vs. temperature phase diagram is located. Anomalous diffusive behavior at low temperature and low bias results in non-monotonic temperature response and negative differential diffusivity; below we clarify how these arise from the inertia of the Brownian particle. At very low bias, a curious temperature-dependent maximum in the forward-biased jump rate is demonstrated, which annihilates a temperature-dependent subcritical minimum in forward-biased jump rate as the temperature increases across this transition. The low-temperature dynamics of the system above the critical bias reveal ringing and second-harmonic generation in the velocity power spectra, hallmarks of stochastic resonant dynamics which demonstrate a convergence of the full dynamics with the zero-temperature limit to complement Cheng and Yip’s analytic work on the average velocity~\cite{ChengYipPhysD2015}.\\
\indent The phase diagram of Fig.~\ref{fig:f3} depicts low- and high-temperature-- equivalently, high- and low-barrier-- driven jump-diffusion regimes. For the unbiased periodic potential, this transition corresponds to the barrier heights below which (or equivalently, temperatures $\delta$ above which) the frequency of the linearized damped oscillations about the shifted well minimum approach the magnitude of the typical inverse time-of-flight for a particle escaping a well at thermal velocity. At lower barriers/higher temperatures, the time-of-flight decreases, the jump rates increase (see Fig.~\ref{fig:f6}), and the probability for multiple jumps increases~\cite{FerrandoPRE1993}. In our scaling, this is consistent with the system approaching the regime of very high damping for all values of forcing~\cite{FerrandoPRE1993,LangBook}\footnote{This holds for temperature-independent friction and well-depth.}, i.e., above about $U_o\sim 6 k_B T$ where, in terms of the Kramers problem, inertial effects become negligible~\cite{FerrandoPRE1993,LangBook}. Above this, the observed low-bias negative differential mobility disappears; the mean forward-biased jumps at zero bias exceed unity and a maximum in the uphill jump appears in the vicinity of the giant diffusion maximum. Our analysis further allows the phase diagram of Speciowicz and Luczka~\cite{SpiechowiczPRE20} to be identified with regimes of persistent subdiffusion at low bias versus a relatively rapid approach to normal diffusion above critical bias, corresponding to the well-known giant diffusion effect~\cite{ReimannPRL2001,ReimannPRE2002}.\\
\indent It is notable that the dynamics of the system are distinct in each region of the $\alpha$-$\beta$ and $\alpha$-$\delta$ parameter spaces of Fig.~\ref{fig:f7} and Fig.~\ref{fig:f3}. For sub-critical $\beta$, the system possesses long-lived ``locked" trajectories, wherein the particles with too low a kinetic energy to cross the barrier remain trapped and oscillate in a single well for a very long time between thermally-activated jumps. At the lowest values of $\beta$ well below the thermal transition, the energy distribution is distorted Maxwellian. The most energetic mode corresponds to the frequency of simple harmonic oscillations about the stretched well minimum, and as seen in Fig.~\ref{fig:f8}, the frequency decreases concurrent with the increased energy distributed to these low frequency modes. Thus the negative differential mobility and decrease to a minimum in jump rate (see also Fig.~\ref{fig:f6}) may be also attributed to inertia; the additional energy input due to increasing bias is effectively stored in the low-frequency simple harmonic oscillator modes corresponding to the linearized, zero-temperature inertial dynamics.\\
\indent Based on the previous discussion of the crossover of the system into the very high damping regime, it appears that the inertia of the Brownian particle is responsible for the persistent subdiffusion,  and hence non-monotonic temperature response for $\beta<\alpha$. This occurs without access to stable ``running" states until $\beta \rightarrow \alpha$, with rare well-to-well and transitions increasing with $\delta$. On the other hand, the increase in $\delta$ beyond about $\sqrt{\alpha/3}$ results an increase in the jump rate (see Fig.~\ref{fig:f6}a), causing more rapid decay of the diffusion coefficient after the initial superdiffusive transient (see Fig.~\ref{fig:f5}). So, while the inertial effects themselves are the main cause of the persistent subdiffusion, the rapid decay is facilitated by the decrease in barrier height relative to temperature leading to increasing multiple jump probability and rate, which itself is a rapidly changing function of friction in the underdamped regime~\cite{FerrandoPRE1993}. It is worth remarking that the friction-bias (or equivalently well depth-bias) and well depth-temperature phase diagrams suggest at least some phenomenological connection between the biased diffusion in a periodic potential and kinetic phase transitions such as glassy behavior and gelation in colloids~\cite{WeitzPRL2001,HoekstraPRE2006}. Below a temperature threshold, there are infinite-waiting time ``locked" states that can only couple thermally to the ``running" states around a threshold bias; at higher temperatures, thermal coupling occurs for any non-zero bias.\\
\indent By considering the velocity power spectra in these various regimes of driven diffusion, the role of the stability of the underlying zero-noise fixed point dynamics becomes clear. The giant diffusion effect corresponds to coupling of trajectories across the I-III bifurcation via thermal fluctuations. Above the bifurcation, characteristic frequencies emerge which show the convergence of full, time-dependent dynamics towards those of the noise-free case. This manifests as transient ``ringing" in the time-dependent diffusivity and non-linear harmonic generation with decreasing temperature evident in the power-spectra for bias above the bifurcation. The evident ringing behavior is of course also reminiscent of inertial systems; as the non-inertial, zero-temperature dynamics do not possess these same oscillatory modes. Overall, this response is consistent with a wide array of stochastic resonant phenomena in periodically-driven systems~\cite{PJungPhysRep93,SchmidPhysicaA05}, and critically, it extends previous analytic results concerning the average velocity~\cite{ChengYipPhysD2015}.\\
\indent Taken together, and contrasted against the overdamped case~\cite{BerezhovskiiDagdugJCP2019}, our analysis suggests inertial effects play a crucial role in the persistent subdiffusion, negative differential mobility, low-bias non-monotonic temperature dependent diffusivity, and even in the specifics of the stochastic resonant dynamics of driven diffusive systems. It is usually taken for granted that inertial effects for diffusion in a potential can be neglected at reasonably long times, or equivalently, when the potential doesn't vary appreciably over the characteristic Kramers diffusion length~\cite{LangBook}. However, in cases of periodic forcing, not only is it feasible that potentials may indeed vary greatly over such a typical length, but moreover any such steep variation occurs twice per period, contributing non-negligibly to the long-time averages.\\
\indent As stated previously the problem of a Brownian particle in a titled washboard potential provides a dynamic model for many non-equilibrium statistical systems. One particularly well-suited system for a driven Brownian particle is that of an active colloid in a periodic optical array. In principle, with the tunability of optical traps, it should be possible to achieve $U_o\ge32k_BT$ to observe the transient ringing in the particle mean-square displacement. While this may or may not be observable in a practical colloid-in-liquid experiment, colloids in gasses or dusty colloidal plasmas~\cite{LowenPRL2019, Lowen2020, Petrov2021} would be an ideal platform. Such an experiment could broadly test the correspondence between the zero-noise dynamics and the high-barrier dynamics. Beyond that, some of the phenomena elucidated herein may find applications in fields such as colloidal assembly or in understanding collective effects in active colloids, where inertial effects have begun to attract attention~\cite{LowenPRL2019, Lowen2020, Petrov2021, GaneshPRE2020}. From a theoretical perspective, while we have considered the velocity dynamics near the I-III bifurcation, the nature of the velocity dynamics near the I-II and II-III may also yield interesting behavior, and the origin of the curious maximum in the forward-biased jump rate at low bias warrants further study, possibly using analytic approaches similar to those of Melnikov~\cite{Melnikov, LangBook}.\\

%\indent  The phenomenon of diffusion enhancement has been verified experimentally using colloids and optical trapping \cite{VolpeAmJPhy2013, VolpePetrovPRE2008, EvstigneevReimannPRE2008}. \\
%
%identification of important parameter regimes for specific physical systems in order to avoid or exploit effects such as giant diffusion enhancement\cite{}, locked vs. running transport regimes \cite{}, and negative differential mobility \cite{BerezhovskiiDagdugJCP2019}, which we have here demonstrated to be a general phenomenon when the energy of the periodic potential dominates the transport response over the characteristic thermal energy, i.e. a lattice-dominant regime. This lattice-dominant regime corresponds to a prevalence of locked-trajectory solutions at forcing sufficiently below the critical forcing, running-trajectory solutions sufficiently above, and the diffusion enhancement effect defining the transition with the temperature-dependent maximum occurring for the critical forcing. This stands in contrast to much simpler bias-dependence in a regime of (lower) diffusion enhancement which grows monotonically towards saturation at large forcing, i.e. a fluctuation-dominant regime, corresponding to running-trajectory solutions occurring with greater frequency at lower forcing.
%Other applications: discuss relevance to active colloids assembly and swarming?\\
\begin{acknowledgments}
This material is based upon work supported by the National Science Foundation under Grant Nos. 2001078 and 2040600 and the work utilized resources from the University of Colorado Boulder Research Computing Group, which is supported by the National Science Foundation (awards ACI-1532235 and ACI-1532236), the University of Colorado Boulder, and Colorado State University.\\
\end{acknowledgments}
\bibliography{bibliography}
\end{document}